\documentclass{ifacconf}
\usepackage[T1]{fontenc}
\usepackage[utf8]{inputenc}
\usepackage{graphicx}      
\usepackage{xcolor}

\usepackage{amsmath, amsfonts, amssymb}
\usepackage{graphicx}
\usepackage{subcaption}
\usepackage{standalone}

\usepackage{tikz}
\usetikzlibrary{intersections}
\usetikzlibrary{positioning}
\usetikzlibrary{arrows,arrows.meta}
\usetikzlibrary{external}
\newcommand{\Norm}[1]{\left\| {#1} \right\|}
\newcommand{\Set}[1]{\left\{ {#1} \right\}}
\newcommand{\E}{{\mathbf E}}
\newcommand{\Reals}{{\mathbb R}}
\newcommand{\Naturals}{{\mathbb N}}
\newcommand{\Hyp}{{\mathcal H}}
\newcommand{\Control}{{\mathcal U}}
\newcommand{\Controls}{{\mathbb U}}

\newcommand{\State}{{\mathcal X}}
\newcommand{\Time}{{\mathcal T}}
\newcommand{\Flow}{{\varphi}}
\newcommand{\Hidden}{{\mathcal Z}}
\newcommand{\Encoder}{{h_{\mathrm{enc}}}}
\newcommand{\Decoder}{{h_{\mathrm{dec}}}}
\newcommand{\Rnn}{{h_{\textrm{RNN}}}}

\usepackage{natbib}        

\begin{document}
\begin{frontmatter}

\title{Learning Flow Functions from Data with Applications to Nonlinear Oscillators\thanksref{footnoteinfo} }
\newcommand*{\email}[1]{\textit{#1}}
\author{Miguel Aguiar$^{1}$,} 
\author{Amritam Das$^{2}$ and} 
\author{Karl H. Johansson$^{1}$}
\thanks[footnoteinfo]{
    © 2023 the authors.
    This work has been accepted to IFAC for publication under a Creative Commons Licence CC-BY-NC-ND.
    An early version of this work was accepted for (non-archival) poster presentation to the NeurIPS 2022 workshop~\textit{The Symbiosis of Deep Learning and Differential Equations}.
}

\address{
    ${}^1${Division~of~Decision~and~Control~Systems}
    and {Digital~Futures},
    KTH~Royal~Institute~of~Technology,
    Stockholm, Sweden and
    ${}^2${Control~Systems~Group, Dept. of Electrical Engineering, Eindhoven University of Technology, Eindhoven, Netherlands.}
}
\vspace{-1ex}
\email{\texttt{\{aguiar,kallej\}@kth.se} and \texttt{am.das@tue.nl}}
\begin{abstract}                
%
We describe a recurrent neural network (RNN) based architecture to learn
the flow function of a causal, time-invariant and continuous-time control system from trajectory data.
By restricting the class of control inputs to piecewise constant functions,
we show that learning the flow function is equivalent to learning the input-to-state map
of a discrete-time dynamical system.
This motivates the use of an RNN together with encoder and decoder networks
which map the state of the system to the hidden state of the RNN and back.
We show that the proposed architecture is able to approximate
the flow function by exploiting the system's causality and time-invariance.
The output of the learned flow function model can be queried at any time instant.
We experimentally validate the proposed method using models of the Van~der~Pol and FitzHugh-Nagumo oscillators.
In both cases, the results demonstrate that the architecture is able to closely reproduce
the trajectories of these two systems.
For the Van der Pol oscillator, we further show that the trained model
generalises to the system’s response with a prolonged prediction time horizon as well as control inputs outside the training distribution.
For the FitzHugh-Nagumo oscillator, we show that the model accurately captures the input-dependent phenomena of excitability.


    
\end{abstract}


\end{frontmatter}

\section{Introduction}
%

Models play a vital role in designing control systems.
For instance, in receding horizon control (\citet{Mac:02}), the model is used to predict the future evolution of the state variables and acts as a constraint in the formulation of the optimal control problem.
With increasing complexity, the curse of dimensionality limits the usefulness of standard first-principle models.
This limitation has motivated research on data-driven approximation of such physical models.
Besides fast predictions for arbitrary initial conditions, another advantage of many data-driven methods in designing control systems is the efficient computation of the gradients of the model with respect to initial conditions, parameters or input signals.
To leverage these advantages, in \citep{Li2021,Geneva2022,Lu2021,Kissas2022}, the focus is on learning the map from initial conditions, parameters and inputs to the solution of a differential equation.

The problem of approximating physical models from data is also tackled in system identification,
see~\citet{Schoukens2019} for an overview.
\citet{Forgione2021}~have proposed methods for identifying dynamics of continuous-time
control systems using neural Ordinary Differential Equations (ODEs). 
However, as the dynamics correspond to the time derivative of the flow,
the neural ODEs must be integrated through an ODE~solver to obtain the system trajectories,
representing an extra computational burden both for prediction and for computing gradients.
Furthermore, errors in the learned dynamics will accumulate over time
when the dynamics are integrated, and the error in the simulated
trajectory can become unbounded.

To directly learn the flow function of an autonomous dynamical system,
\citet{Bilos2021}~has proposed an alternative to neural~ODEs that avoids the step of using an ODE solver and
allows for faster prediction.
This motivates the search for a corresponding learning scheme for controlled dynamical systems where inputs are present.
However, this is a harder problem since the domain of the flow of a
continuous-time control system is infinite-dimensional.

\citet{Hanson2020}~have shown that continuous-time recurrent neural networks are universal 
approximators for flow functions of stable continuous-time dynamical systems,
where the approximation quality is uniform over time.
But the question of whether learning such a model
from data is feasible in practice is to the best of our knowledge open.

Our main contributions are a neural network-based architecture for 
learning flow functions of controlled dynamical systems in continuous-time and
a detailed experimental demonstration of the performance and generalisation capabilities of 
the proposed architecture in predicting the responses of nonlinear oscillators.
We provide a mathematical formulation of the problem of learning the flow function of a control system,
showing that it can be reduced to a tractable optimisation problem.
The inputs are restricted to the class of piecewise constant functions,
which are practically relevant since digital controllers typically produce such control inputs.
Leveraging the causality and time-invariance properties of the considered class of systems,
we show that the continuous time flow function can be efficiently approximated by a discrete-time recurrent neural network-based architecture.
We demonstrate the capabilities of the proposed architecture in predicting the input-dependent response
of nonlinear oscillators such as the Van der Pol and FitzHugh-Nagumo oscillators.

The organisation of the paper is as follows.
After a detailed description of the considered class of control systems and formulating the learning problem in Section~\ref{sec:formulation},
in Section~\ref{sec:architecture} we present the proposed architecture to learn the flow function of a control system.
In Section~\ref{sec:van-der-pol}, the proposed methodology is experimentally evaluated to predict
the periodic response of the non-autonomous Van der Pol oscillator.
In Section~\ref{sec:fitzhugh-nagumo}, the proposed methodology is evaluated to predict
the occurrence of excitable behaviour in the FitzHugh-Nagumo oscillator.
Finally, in Section~\ref{sec:conclusion} we provide some concluding remarks and future directions for further research. 

\section{Problem formulation}\label{sec:formulation}
\subsection{Considered class of control systems and flow functions}
A control system $\Sigma$ consists of the following quadruple (see \citet{Sontag1998}, pp. 26)
\begin{align}
\label{eq:dynamics}
   \Sigma =  (\Time, \State, \Controls, \Flow),
\end{align}
describing the evolution of state-variables of the dynamical system over a time interval $\mathcal{T}$
depending on its initial condition $x \in \State$ and input $u \in \Controls$,
where $\Controls$ is a set of functions $u: \Time \to \Control$.
The \emph{flow}, dictating this evolution, is defined as a mapping $\Flow: \Time \times \State \times \Controls \rightarrow \State$. 

We assume that $\Sigma$ is time-invariant and finite dimensional. In particular, $\Time \subseteq \mathbb{R}_{\geq 0}$, $\State \subset \Reals^n $ and $\Control \subset \Reals^m$. We also assume that $\Controls$ is the set of piecewise constant controls%
\footnote{
    The approach can be generalised to any class of input signals that admit a finite-dimensional causal parameterisation.
} of period $\Delta>0$. In other words, given a sequence $\Set{u_k}_{k=1}^\infty$ with $u_k \in \Control$, the control input $u$ is defined by
\begin{equation}
\label{input}
    u(t) = u_k, \ (k - 1)\Delta \leq t < k\Delta, \ k \in  \Naturals.
\end{equation}
We will exploit two properties of the flow: \emph{causality} which implies that the flow at time $T\geq 0$, $\Flow(T, x, u)$ depends only on the values of $u(t)$ for $0 \leq t < T$,
and \emph{continuity} in the sense that $t \mapsto \Flow(t, x, u)$ is continuous for each $x, u$.

As an example, we can consider the flow function generated by a system of ordinary differential equations ${\dot{\xi}(t) = f(\xi(t), u(t))}, \ t \geq 0$ with initial condition $x$ where $u$ is generated by a digital controller.

\subsection{Mathematical formulation of the learning problem}
We are interested in learning the flow from data
on a time interval $[0, T]$ with $T > \Delta$.
The input signals $u$ and initial conditions $x$ of interest are assumed to be drawn from probability distributions $P_u$ on $\Controls$ and $P_x$ on $\State$, respectively.

Given an hypothesis class $\Hyp \subset \Set{\hat\Flow: \Reals_{\geq 0} \times \State \times \Controls \to \State}$, we define for $\hat\Flow \in \Hyp$ the loss function
\begin{equation}\label{eq:lossfn}
    \ell_T(\hat\Flow) :=
    \E\left[ { \frac{1}{T} \int_{0}^T{ \Norm{ \hat\Flow(t, X, U) - \Flow(t, X, U) }^2\mathrm{d}t}} \right],
\end{equation}
(the average squared prediction error over $[0, T]$),
where $X \sim P_x$ and $U \sim P_u$ are independent random variables.
The problem of finding the best approximation to the true flow $\Flow$
amounts to minimising $\ell_T$ over $\Hyp$.

In practice, the data consists of discrete-time samples from $N$ different trajectories:
\begin{equation}\label{eq:measurements}
    \xi^i_k = \Flow(t^i_k, x^i, u^i) + v^i_k, \ k = 1, \dots, K, \ i = 1, \dots, N,
\end{equation}
where $K$ is the number of samples of each trajectory,
$t^i_k \in [0, T]$ is an increasing sequence of time samples,
$v^i_k$ is measurement noise,
and $x^i, u^i$ are sampled i.i.d. from $P_x$ and $P_u$.
Thus, in order to learn $\Flow$ from the data~\eqref{eq:measurements},
we define the empirical loss function 
\begin{equation}
    \hat\ell_T(\hat{\Flow}) := \frac{1}{N}\sum_{i = 1}^N{
        \frac{1}{K}\sum_{k = 1}^{K} \Norm{\xi^i_k - \hat\Flow(t^i_k, x^i, u^i)}^2
    }
\end{equation}
and search for a minimiser of $\hat\ell_T$ in $\Hyp$.

The objective of this paper is to define an hypothesis space $\Hyp$ which renders the above problem tractable and
provides an approximation $\hat\Flow$ of the true flow function $\Flow$ while preserving causality and continuity.
In the next section, we propose a neural network-based architecture to solve this problem.

\section{Proposed architecture}\label{sec:architecture}

\begin{figure*}[htbp]
    \centering
    \begin{subfigure}{0.49\textwidth}
        \centering
        \includegraphics[width=0.99\textwidth]{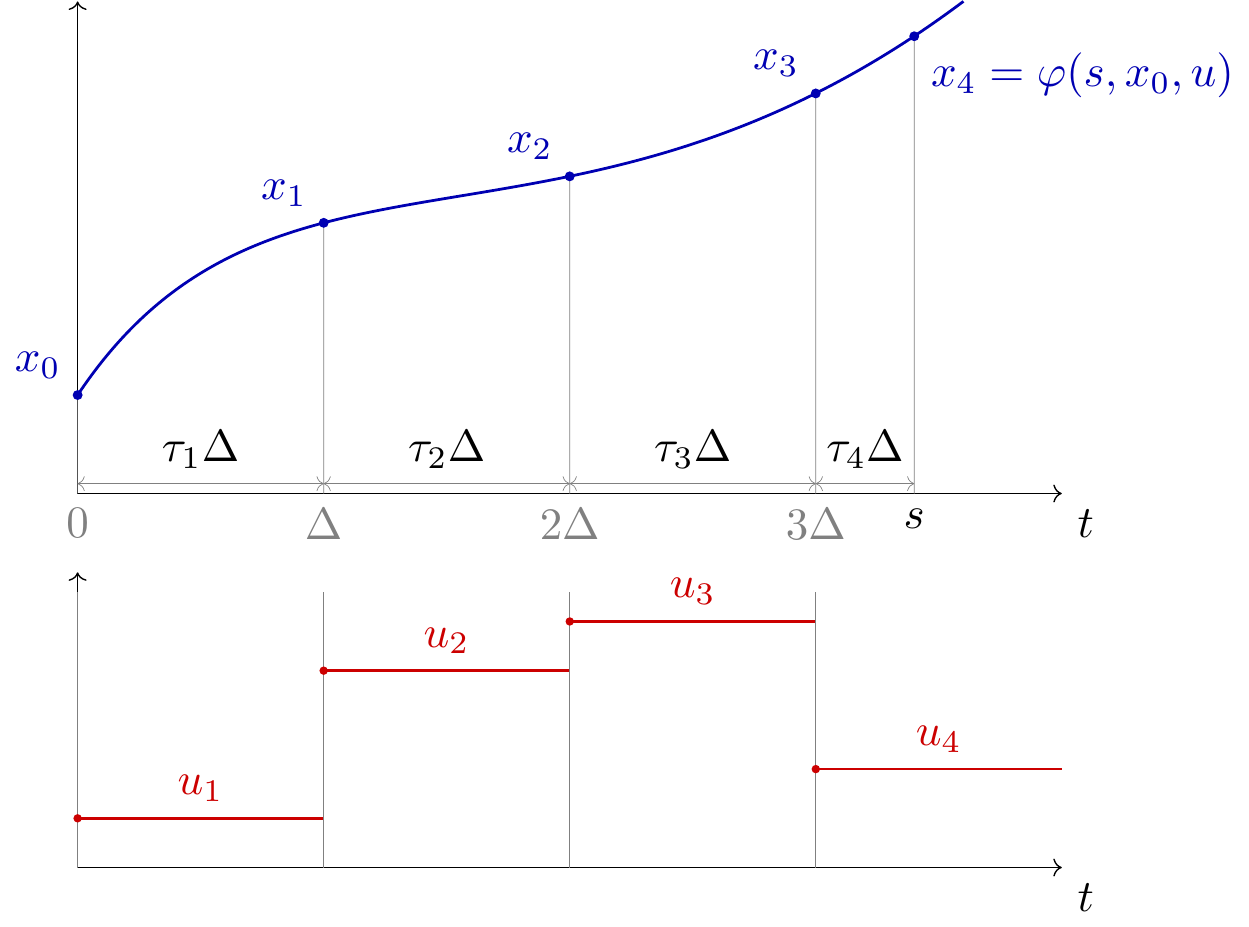}
        \vspace{-1em}
        \subcaption{}
        \label{fig:nn-arch-plot}
    \end{subfigure}
    \begin{subfigure}{0.49\textwidth}
        \centering
        \includegraphics[width=0.99\textwidth]{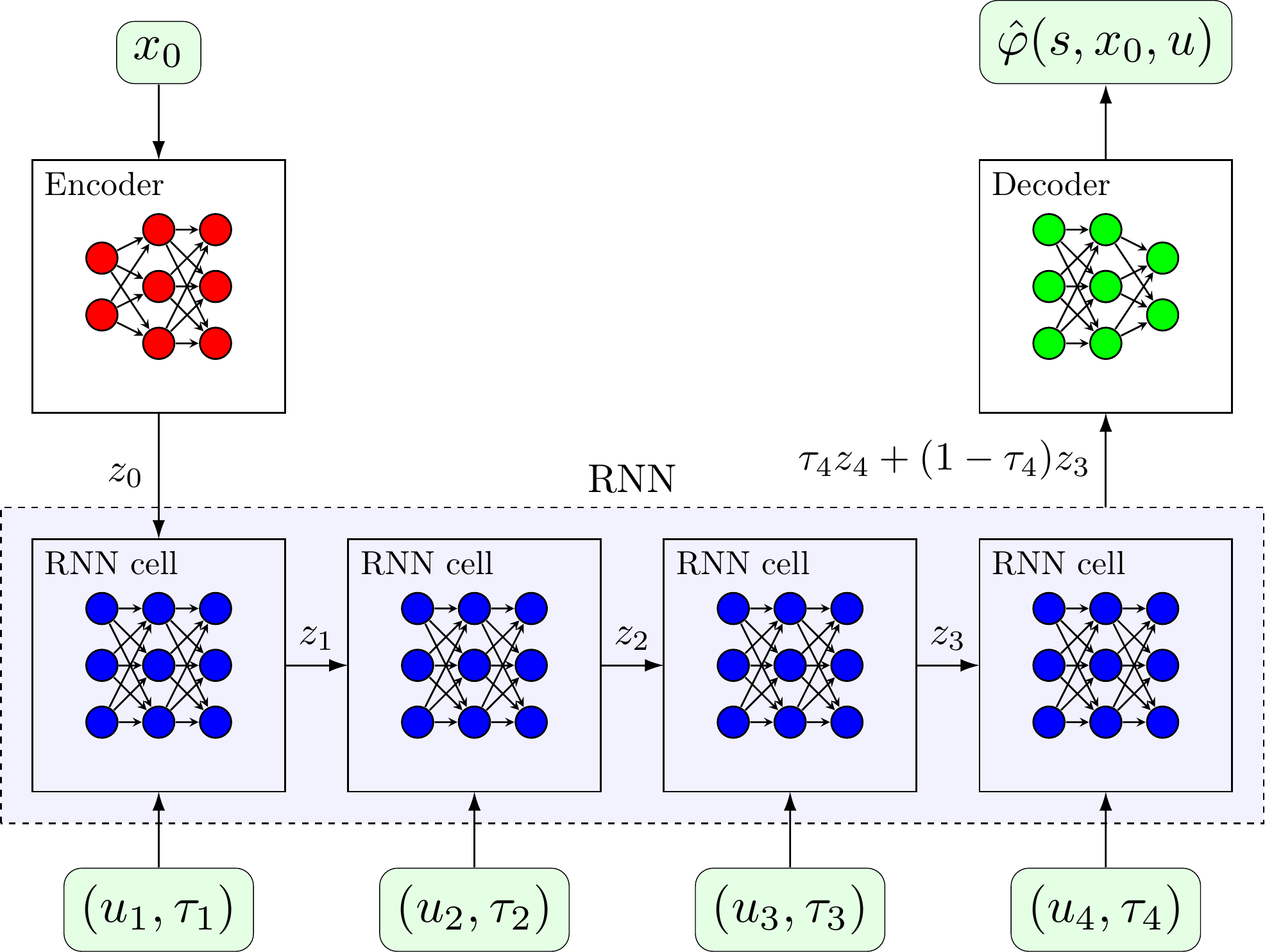}
        \subcaption{}
        \label{fig:nn-arch-blocks}
    \end{subfigure}
    \vspace{-0.5em}
    \caption{
        (a) Schematic illustration of true flow function $\Flow$ for parameters $\{u_k, \tau_k\}_{k=1}^4$.
        (b) Corresponding model for the approximated flow $\hat{\Flow}$.
        In the approximated model, we first map the initial condition to a higher dimensional space through a feedforward \emph{encoder} network.
        Then, the encoded state is propagated in time through a Recurrent Neural Network~(RNN).
        Each cell of the RNN sequentially takes $(u_k, \tau_k)$ as inputs.
        The two last hidden states are interpolated and mapped back to $\State$ through another feedforward \emph{decoder} network.
    }
    \label{fig:nn-arch}
      \hrulefill
\end{figure*}

\subsection{Motivation}
Due to causality and the considered class of inputs \eqref{input},
the flow $\Flow(s, x, u)$ at a time instant $s \in \Reals_{\geq 0}$ depends only on a finite number of the input values $\Set{u_k}$.
Thus, at any time during the first control period $[0, \Delta]$,
only the value of $u_1 \in \Control$ and the initial condition $x \in \State$ are required to define $\Flow$.
Therefore, we define $\Phi: [0, 1] \times \State \times \Control \rightarrow \State$ as
\begin{equation*}
   \Phi(\tau, x, u_1) := \Flow(\tau\Delta, x, u),
\end{equation*}
such that a finite-dimensional vector of parameters (as opposed to functions) directly maps to the flow.
For an arbitrary time instant $s \in \Reals_{\geq 0}$, the flow $\Flow(s, x, u)$ can be computed as follows:
\begin{enumerate}
\item Construct a map $\mathrm{d}_{\Delta}: \left(s, u\right) \mapsto \{\tau_k, u_k\}_{k=1}^{k_{s}+1}$ such that $u_k$ is given according to \eqref{input} and 
\begin{equation*}
   k_s := \left\lfloor \frac{s}{\Delta} \right\rfloor, \quad\quad \tau_k := \begin{cases}
        {1}, & {k \leq k_s} \\
        \dfrac{s - {k_s}\Delta}{\Delta}, & {k = k_s + 1}.
    \end{cases}
\end{equation*}
\item Define the sequence ${x_k} \in \State$ for all $k = 1, \dots, k_s + 1$ as
\begin{align}
\label{eq:discrete-equivalent}
    x_0 &= x, \notag\\
    x_{k} &= \Phi(\tau_k, x_{k - 1}, u_k).
\end{align}
\end{enumerate}
Then we have that $x_{k_s + 1} = \Flow(s, x, u)$.
Thus, trajectories of $\Flow$ can be equivalently represented by the trajectories of a discrete-time dynamical system with inputs $(\tau_k, u_k)$, as represented in \eqref{eq:discrete-equivalent} and illustrated in Figure~\ref{fig:nn-arch-plot}.

\subsection{Definition of the architecture}
Based on the previous discussion, we focus on approximating
the set of difference equations~\eqref{eq:discrete-equivalent},
and Recurrent Neural Network (RNN) models are suitable for this task
since they are universal approximators of such mappings~\citep{Schafer2006}.


To increase the flexibility of the model, we first map the initial state to a feature space~$\Hidden$
using an \emph{encoder} network, a deep neural network (DNN) whose map we denote by $\Encoder$.
Denoting the mapping defined by the RNN as $\Rnn$, we then have
\begin{equation*}
\begin{aligned}
    z_0 &= \Encoder(x), \\
    (z_1, \dots, z_{k_s + 1}) &= {\Rnn}{\left( z_0, \Set{(u_k, \tau_k)}_{k=1}^{k_s + 1} \right)},
\end{aligned}
\end{equation*}
where $z_i \in \Hidden$ are the hidden states of the RNN.
To ensure that the learned flow function is continuous in time,
we combine the two last RNN states using the following map:
\begin{equation*}
    g{\left(\Set{z_k}_{k=1}^{k_s+1}, \Set{\tau_k}_{k=1}^{k_s + 1}\right)} := {(1 - \tau_{k_s + 1})}z_{k_s} + \tau_{k_s + 1}z_{k_s + 1}.
\end{equation*}
Note that this does not amount to linear interpolation since $z_{k_s + 1}$ depends on $\tau_{k_s + 1}$.
To map the output of $g$ back to a state vector in $\State$ we use a \emph{decoder} DNN which we denote by $\Decoder$,
whose output yields the approximated flow $\hat\Flow$ at the time instant $s$, i.e.
\begin{equation}
\label{eq:hdec}
   \hat\Flow(s, x, u) = {\Decoder}{\left( z \right)},
\end{equation}
where $z$ is defined as
\begin{equation}
\label{eq:z}
    z := g{\left(
        \Rnn{\left(\Encoder(x), \Set{\tau_k, u_k}_{k=1}^{k_s + 1} \right) },
        \Set{\tau_k}_{k=1}^{k_s + 1}
    \right)}.
\end{equation}
As a result, the hypothesis space $\Hyp$ is defined by the set of functions $\hat\Flow$ of the form \eqref{eq:hdec}-\eqref{eq:z} that are
parameterised by the parameters of the networks $\Rnn, \Encoder$ and $\Decoder$.

A block diagram of the architecture is shown in Figure~\ref{fig:nn-arch-blocks}.

\section{Experimental evaluation: Van der Pol Oscillator}\label{sec:van-der-pol}


\begin{figure*}[htbp]
    \centering
    \begin{subfigure}{.49\textwidth}
        \centering
        \includegraphics[width=.8\linewidth]{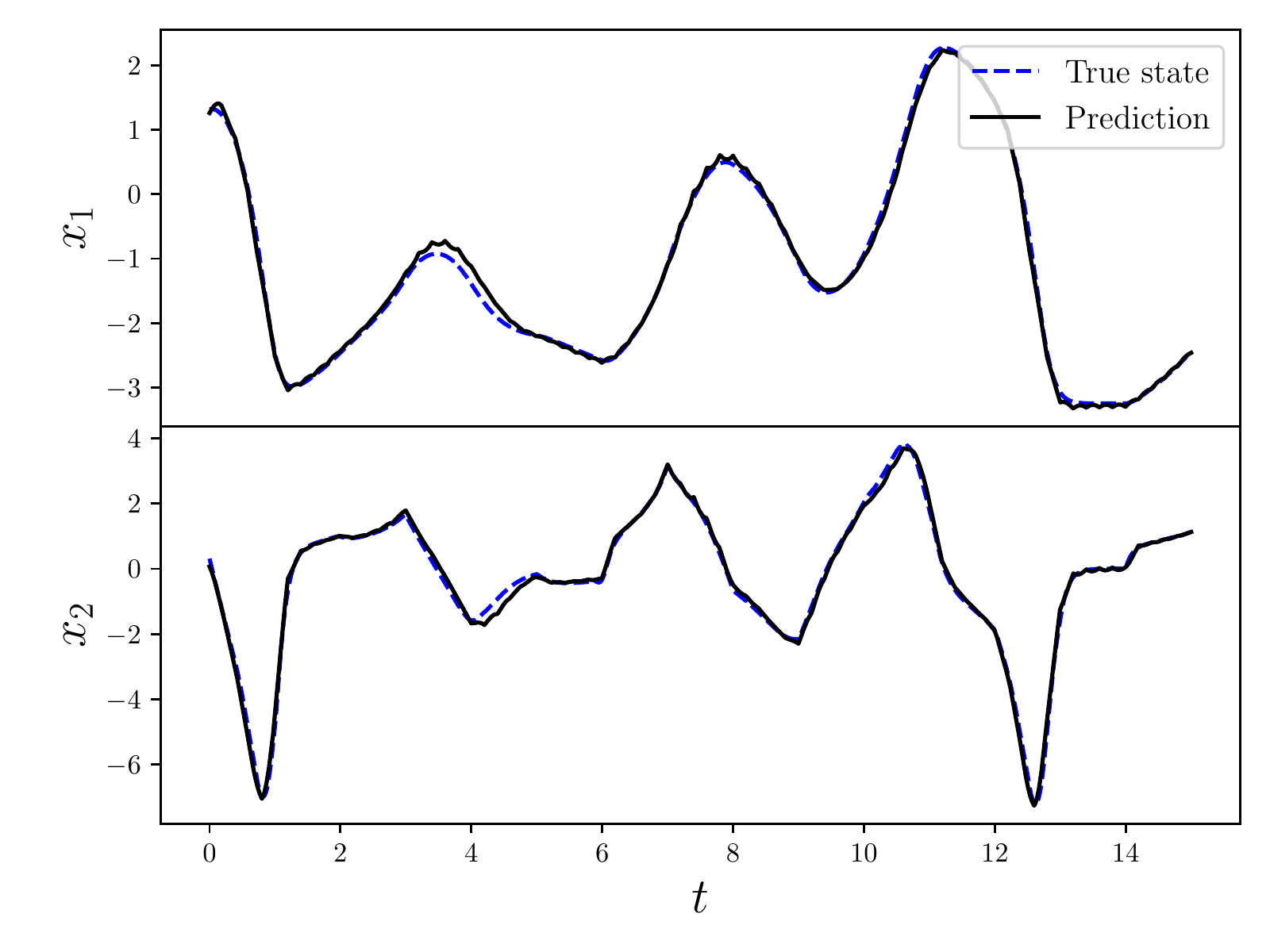}  
    \end{subfigure}
    \begin{subfigure}{.49\textwidth}
        \centering
        \includegraphics[width=.8\linewidth]{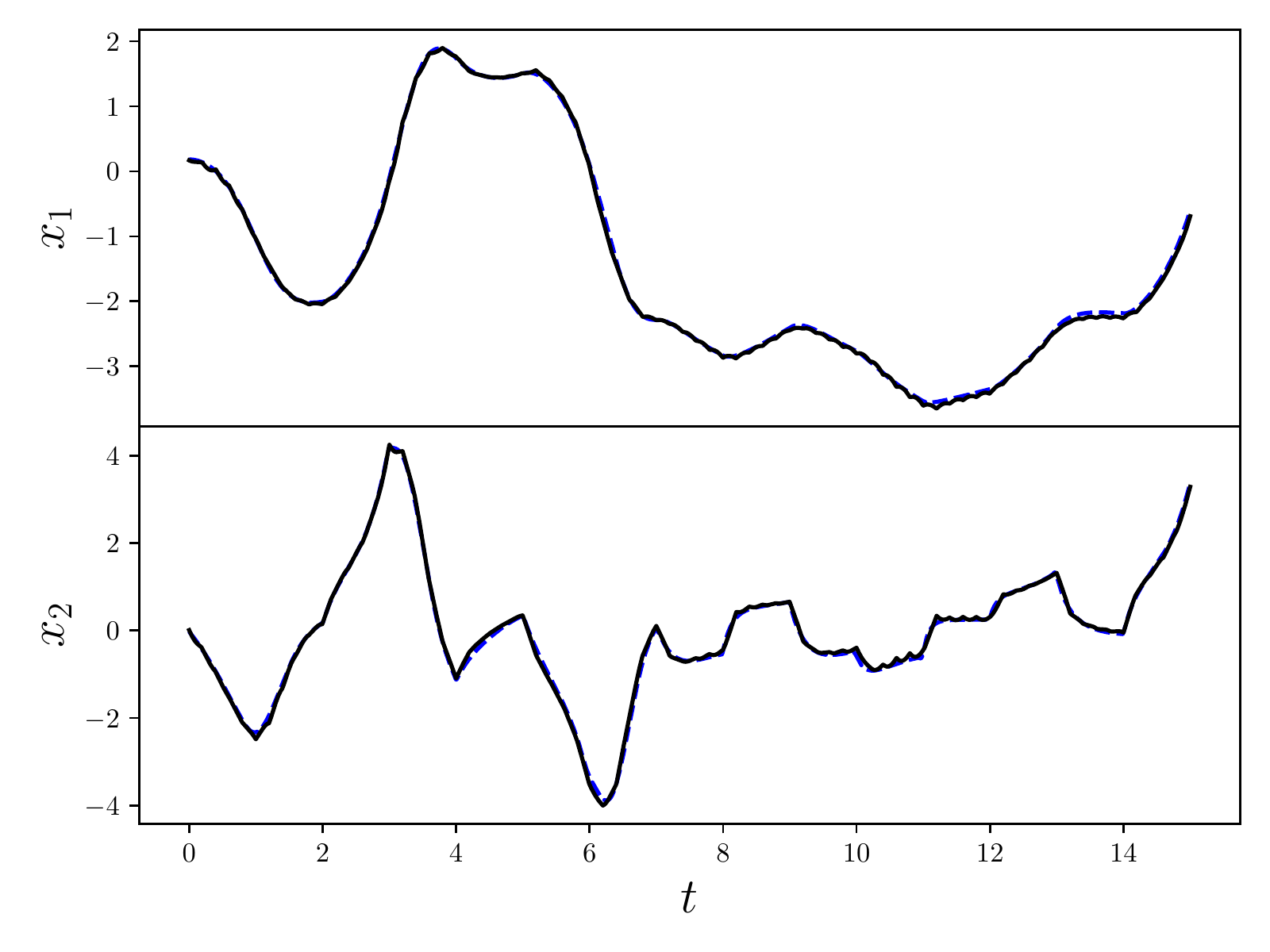} 
    \end{subfigure}
    \vspace{-0.75em}
    \caption{
        Actual (blue, dashed) and predicted (black) trajectories for the Van der Pol model with initial condition
        and input drawn from the corresponding distributions.
        }
    \label{fig:vdp-prediction}
        \hrulefill
\end{figure*}

We illustrate the proposed method by evaluating its performance in 
learning the flow function of the Van der Pol oscillator. 
Additionally, we study the generalisation capabilities of the trained model
with respect to the simulation time horizon and the input distribution.

\subsection{Learning the flow function}
\subsubsection{Data generation:}
The Van der Pol oscillator is described by the system of ordinary differential equations
\begin{equation}\label{eq:vdp-dynamics}
\begin{aligned}
    \dot{x}_1(t) &= x_2(t) \\
    \dot{x}_2(t) &= -x_1(t) + \left(1 - x_1(t)^2\right)\mu x_2(t) + u(t).
\end{aligned}
\end{equation}
We take $\mu = 1$ and $x(0) \sim N(0, I)$, i.e. a standard normal distribution.
The control input sampling time is $\Delta = 0.2$ and the inputs considered are
square wave inputs with period $5\Delta$ and amplitudes sampled i.i.d. from ${N(0, \sigma = 5)}$, i.e.
\begin{equation*}
\begin{aligned}
    u_{1 + 5k} &\sim N(0, 5),\\
    u_{j + 5k} &= u_{1 + 5k}, \ j = 2, 3, 4, 5
\end{aligned}
\end{equation*}
holds for all $k \geq 0$.

To generate the data used to train the model,
we integrate~\eqref{eq:vdp-dynamics} with an RK45~solver.
A total of ${N = 30}$ trajectories are generated,
and these are divided into train, validation and test sets
following a $60\% - 20\% - 20\%$ random split.
For each trajectory, $K = 200$ time points $t^i_k$ are sampled using Latin hypercube sampling.
The measurement noise in~\eqref{eq:measurements} is
zero-mean Gaussian noise with standard deviation of~0.1.

\subsubsection{Training:}
We train the model using the stochastic gradient descent algorithm \texttt{Adam}
with a batch~size of 1024.
The learning rate is reduced 5-fold for every 5~epochs in which the 
validation loss is not reduced,
and the training is stopped when the validation loss does not
decrease more than $5\times 10^{-4}$ for 30~consecutive~epochs.

\subsubsection{Results:}
We used random search to determine the best values for the size of the
encoder, decoder and recurrent networks, and the initial learning rate.
The optimal initial learning rate was found to be $1\times 10^{-2}$
and the training took 186~seconds on a cluster node with an NVIDIA T4 GPU and
an Intel(R) Xeon(R) Gold 6226R CPU @ 2.90GHz.
The RNN is a single-layer LSTM network with 8~hidden states.
The encoder network maps the initial state $x$ to the initial LSTM hidden and cell states,
and is a 3-layer feedforward net with 96 nodes in the hidden layers.
The decoder network maps the hidden LSTM state to the flow value and
is a 3-layer feedforward network with 48 nodes in the hidden layers.
All networks use $\tanh$ activations.

Figure~\ref{fig:vdp-prediction} shows two predicted trajectories on $[0, 15]$ for two new pairs of
initial conditions and inputs drawn from $P_x$ and $P_u$ (i.e., unseen during training).
Note that the actual trajectory and the predicted trajectory are nearly indistinguishable.

\subsection{Prediction over large time horizons}

\begin{figure}[htbp]
    \centering
    \includegraphics[width=.9\linewidth]{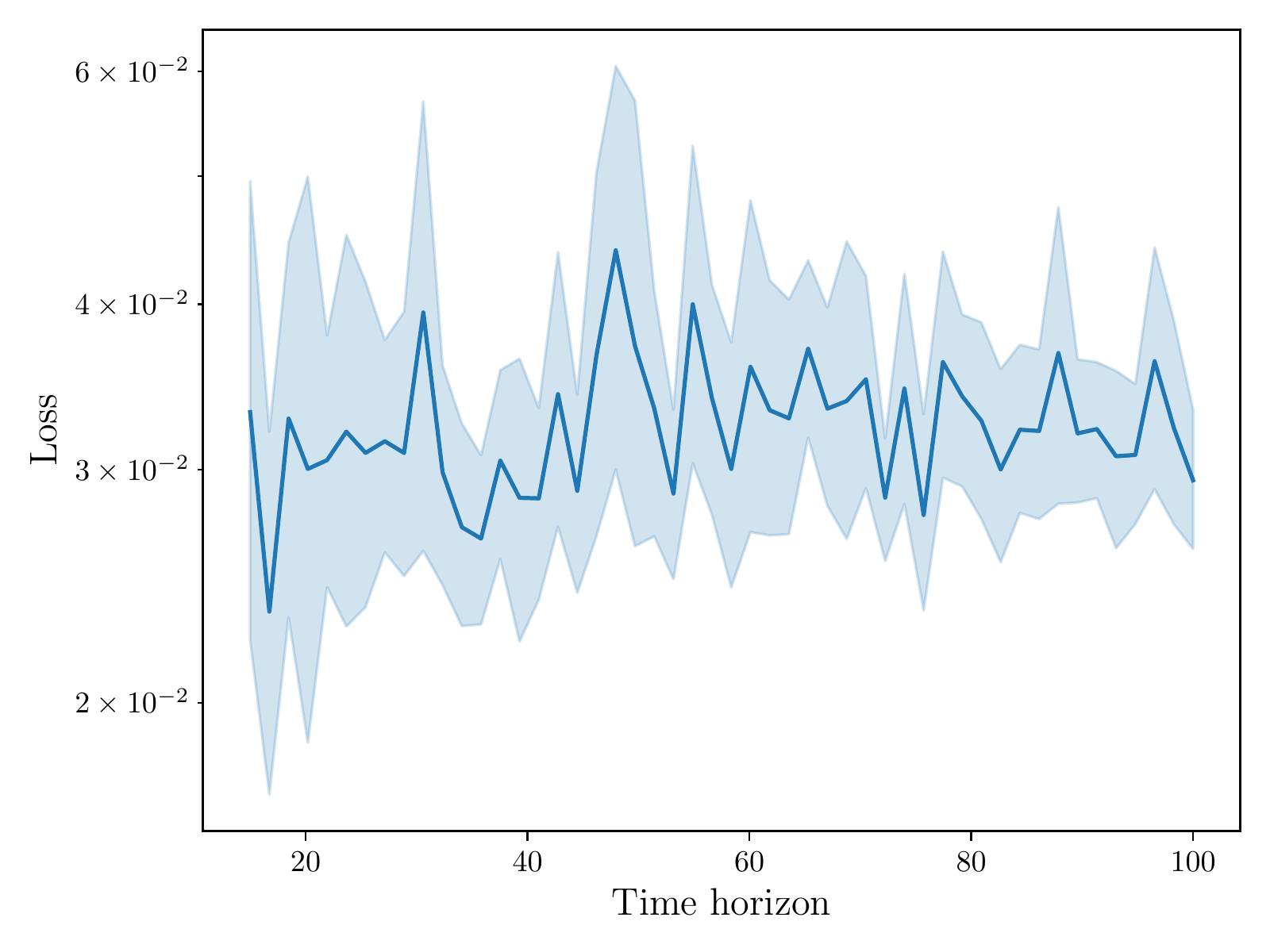} 
    \vspace{-0.5em}
    \caption{
        Estimate of $\ell_t(\hat\Flow)$ as a function of $t$ for the Van der Pol model trained with $T = 15$.
    }
    \label{fig:vdp_loss_time_horizon}
\end{figure}

The trajectories used for training the model of the previous section have a time horizon of $15$~seconds.
Due to the recurrent structure of the model and the stability of the system under consideration,
we expect that the performance is maintained for longer prediction horizons.
To this end, we compute the test loss $\ell_t(\hat\Flow)$
on a different set of $100$ trajectories on $[0, t]$
for a set of gridded values ${15 \leq t \leq 100}$.
The result is shown in Figure~\ref{fig:vdp_loss_time_horizon},
where the coloured area represents the 95\% confidence interval approximated using the empirical variance estimate.
We observe that $\ell_t$ remains approximately constant as $t$ increases,
indicating that the model gives reliable predictions
for $t$ much larger than the value of $T$ used for the training trajectories.

\subsection{Generalisation to different input distributions}\label{app:input-dist}
    
\begin{figure}[htbp]
    \centering
    \includegraphics[width=0.9\linewidth]{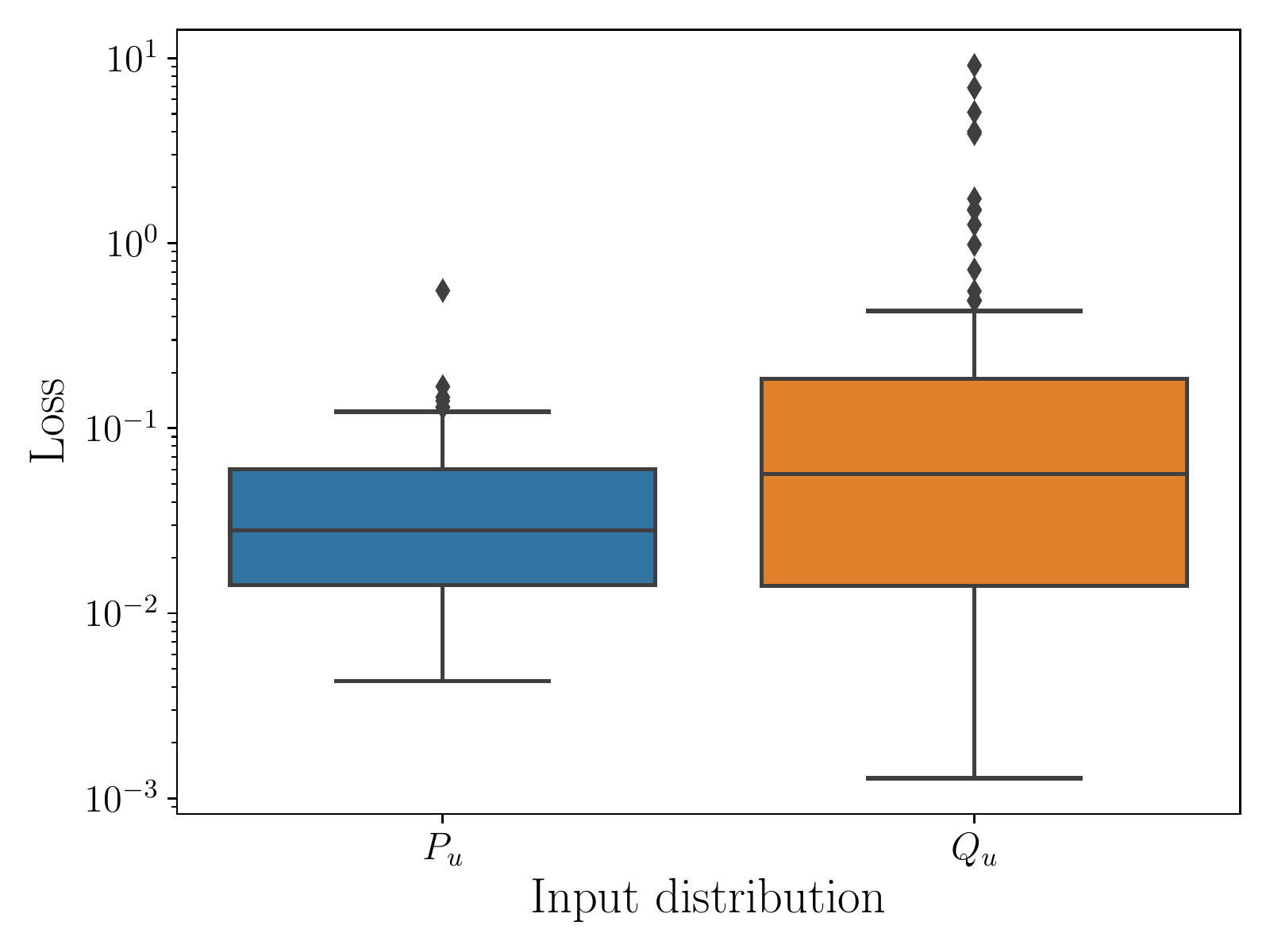}
    \vspace{-0.5em}
    \caption{Distribution of the loss $\ell_T(\hat\Flow)$ with input distributions $P_u$ (blue) and $Q_u$ (orange).}
    \label{fig:app-vdp-loss-sine}
\end{figure}

We additionally investigate the performance of the model trained in section~\ref{sec:van-der-pol}
with an input distribution different from the training distribution $P_u$.
In particular, we consider a new distribution $Q_u$ on $\Controls$ consisting of sinusoidal sequences
with random amplitude and frequency, that is,
\begin{equation*}
    u_k = A\, {\sin}{\left( {\Omega}k\Delta \right)},
\end{equation*}
where ${A\sim\mathrm{LogNormal}(0, 1)}$ and ${\Omega\sim\mathrm{Uniform}(0, 2\pi)}$.
This corresponds to sinusoidal signals with a maximum frequency of $1$~Hz.

To verify the performance on this class of inputs, we compute an estimate of $\ell_T(\hat\Flow; Q_u)$,
defined as in equation~\eqref{eq:lossfn} with the expectation taken with ${U \sim Q_u}$.
Figure~\ref{fig:app-vdp-loss-sine} shows a box plot of the distribution of the estimate of $\ell_T$
computed on 100~trajectories for each of the two input distributions $P_u$ and $Q_u$.
As expected, the mean and variance of the prediction loss for the distribution $Q_u$ are slightly higher
than for $P_u$, but remain reasonably close to that of $P_u$.

\section{Predicting excitability in the FitzHugh-Nagumo Oscillator}\label{sec:fitzhugh-nagumo}

\begin{figure*}[htbp]
    \centering
    \begin{subfigure}{.46\textwidth}
        \centering
        \includegraphics[width=.85\linewidth]{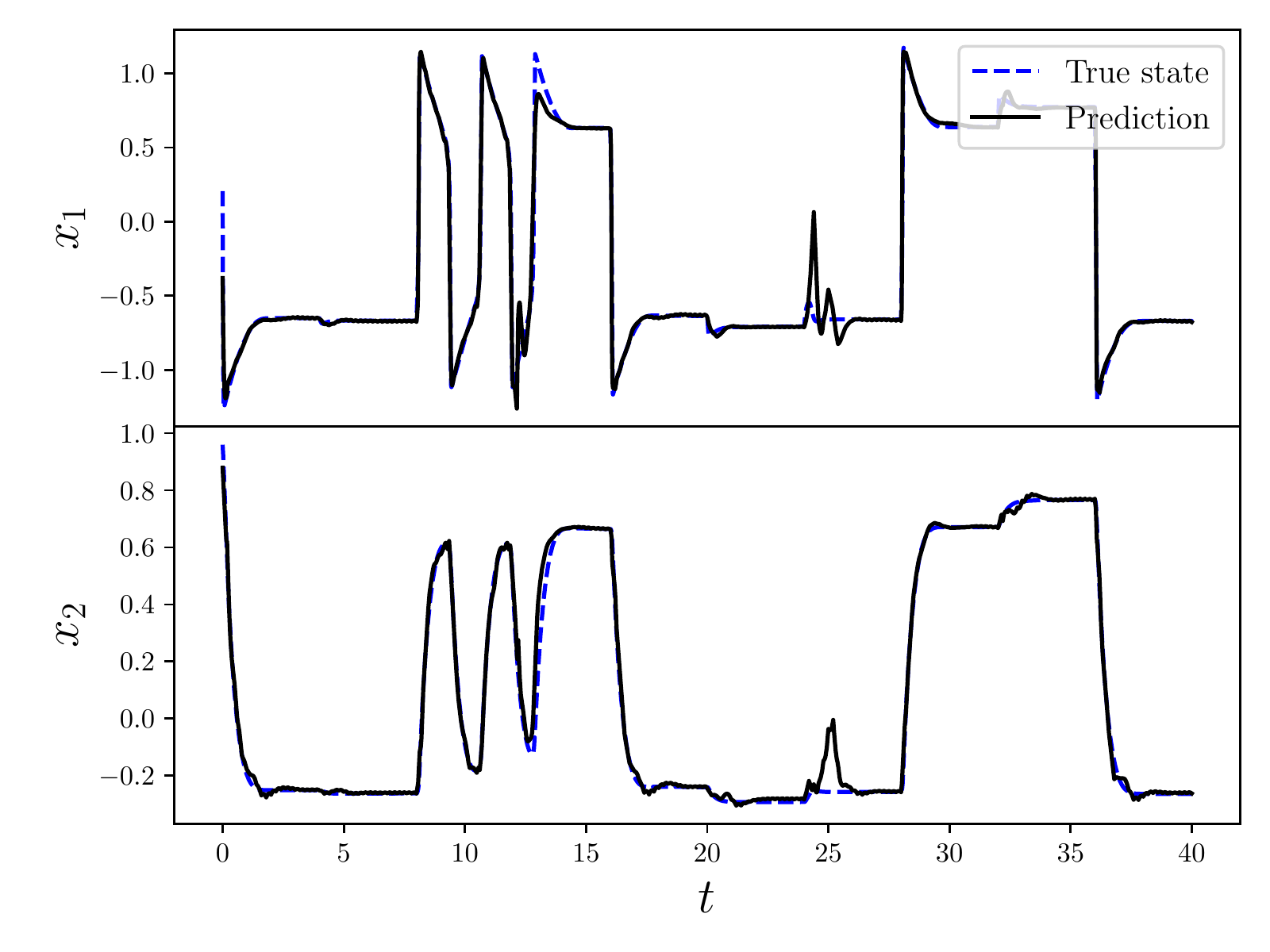}  
        \vspace{-0.7em}
        \subcaption{}
    \end{subfigure}
    \begin{subfigure}{.46\textwidth}
        \centering
        \includegraphics[width=.85\linewidth]{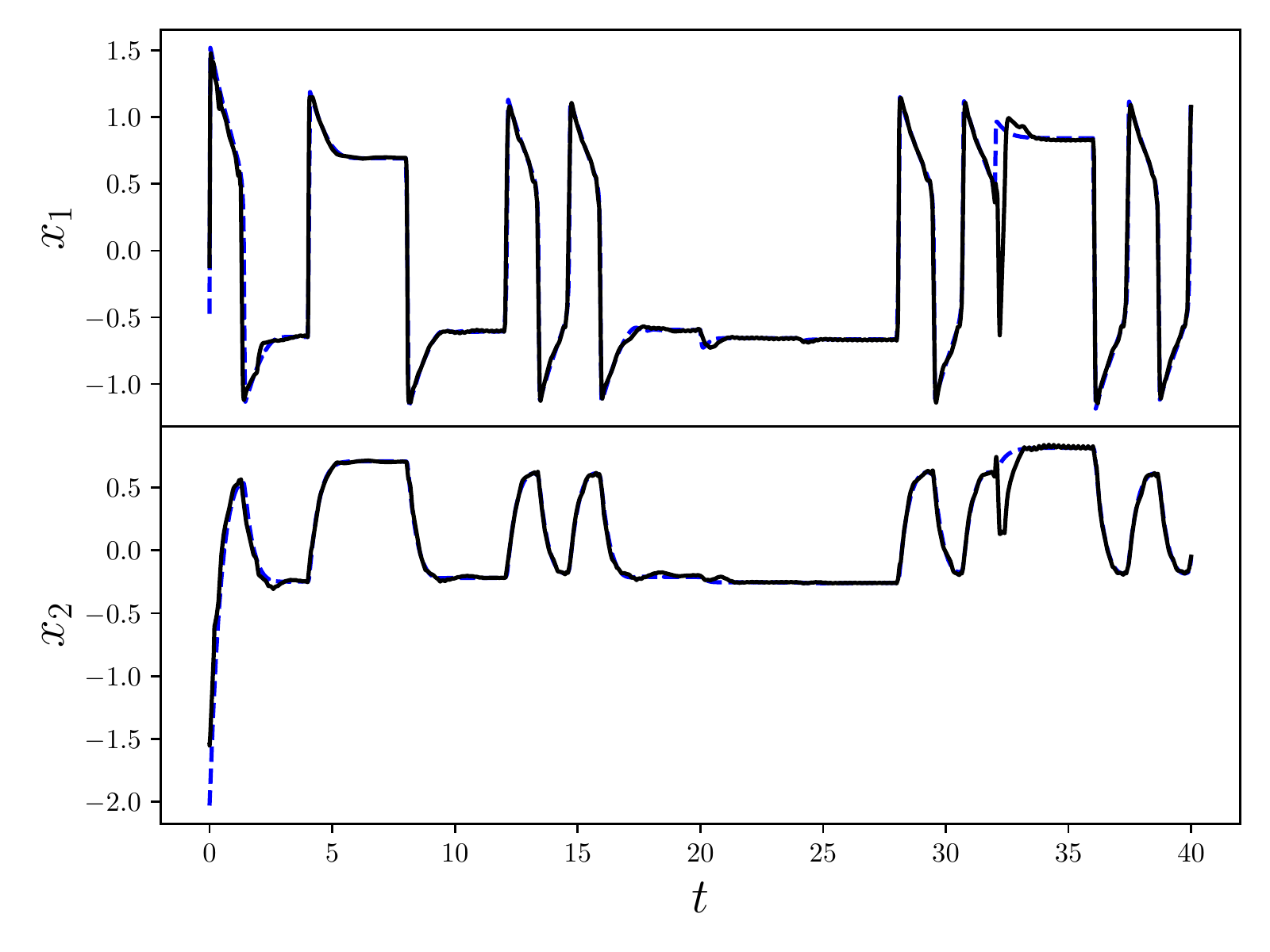} 
        \vspace{-0.7em}
        \subcaption{}
    \end{subfigure}
    \vspace{-0.5em}
    \caption{
     Actual (blue, dashed) and predicted (black) trajectories for the FitzHugh-Nagumo oscillator with initial condition and input drawn from the corresponding distributions.
    }
    \label{fig:fhn-prediction}
    \vspace{-0.5em}
\end{figure*}

Excitability is the system property of biological oscillators that constitute neurons, muscle cells, and endocrine cells.
Here, based on the input energy, the output response either exhibits resting behaviour or stereotypical spike trains (also known as an all-or-none response, see \cite{2017arXiv170404989S}). Predicting the excitable behaviour has been a key problem in neurophysiology.
In this paper, we provide a data-driven method where we apply the proposed model architecture for learning the flow of FitzHugh-Nagumo oscillator and predict its excitable behaviour. 

\subsection{Learning the flow function}
\subsubsection{Data generation:}
The FitzHugh-Nagumo oscillator is described by the following set of nonlinear differential equations:
\begin{align}
\label{eq:fhn-dynamics}
    \eta\dot{x}_1(t) &= x_1(t) - x_1(t)^3 - x_2(t)+ u(t), \notag\\
    \eta\gamma\dot{x}_2(t) &= x_1(t)+ a - bx_2(t),
\end{align}
where $\eta, \gamma, a$ and $b$ are positive constants,
which we choose as $\eta = 1/50, \, \gamma = 40, \, a = 0.3, \, b = 1.4$.

As before, we take $x(0) \sim N(0, I)$.
The control period is $\Delta = 0.1$ and the input distribution $P_u$ is given by
\begin{equation*}
\begin{aligned}
    u_{1 + 40k} &\overset{\text{i.i.d.}}{\sim}\mathrm{LogNormal}(\mu=\log(0.2), \sigma=0.5),\\
    u_{j + 40k} &= u_{1 + 40k}, \ j = 2, \dots, 40
\end{aligned}
\end{equation*}
for all $k \geq 0$.
This is chosen so that the multi-stable behaviour of the oscillator is observed.

We generated $N = 300$ trajectories on $[0, 20]$ using an RK45 solver, sampling $K = 300$
time points from each trajectory using Latin hypercube sampling.
The measurement noise in~\eqref{eq:measurements} is
zero-mean Gaussian noise with standard deviation equal to $0.05$.

\subsubsection{Results:}
We trained a model where the RNN is an LSTM with 16 hidden states,
the encoder has 2 hidden layers with 64 nodes each,
and the decoder has 2 hidden layers with 32 nodes each.
The training took 1196~seconds with an initial learning rate of $2\times10^{-2}$,
with the same algorithm and hardware as in Section~\ref{sec:van-der-pol}.

\begin{figure*}[htbp]
    \centering
    \begin{subfigure}{.46\textwidth}
        \centering
        \includegraphics[width=.85\linewidth]{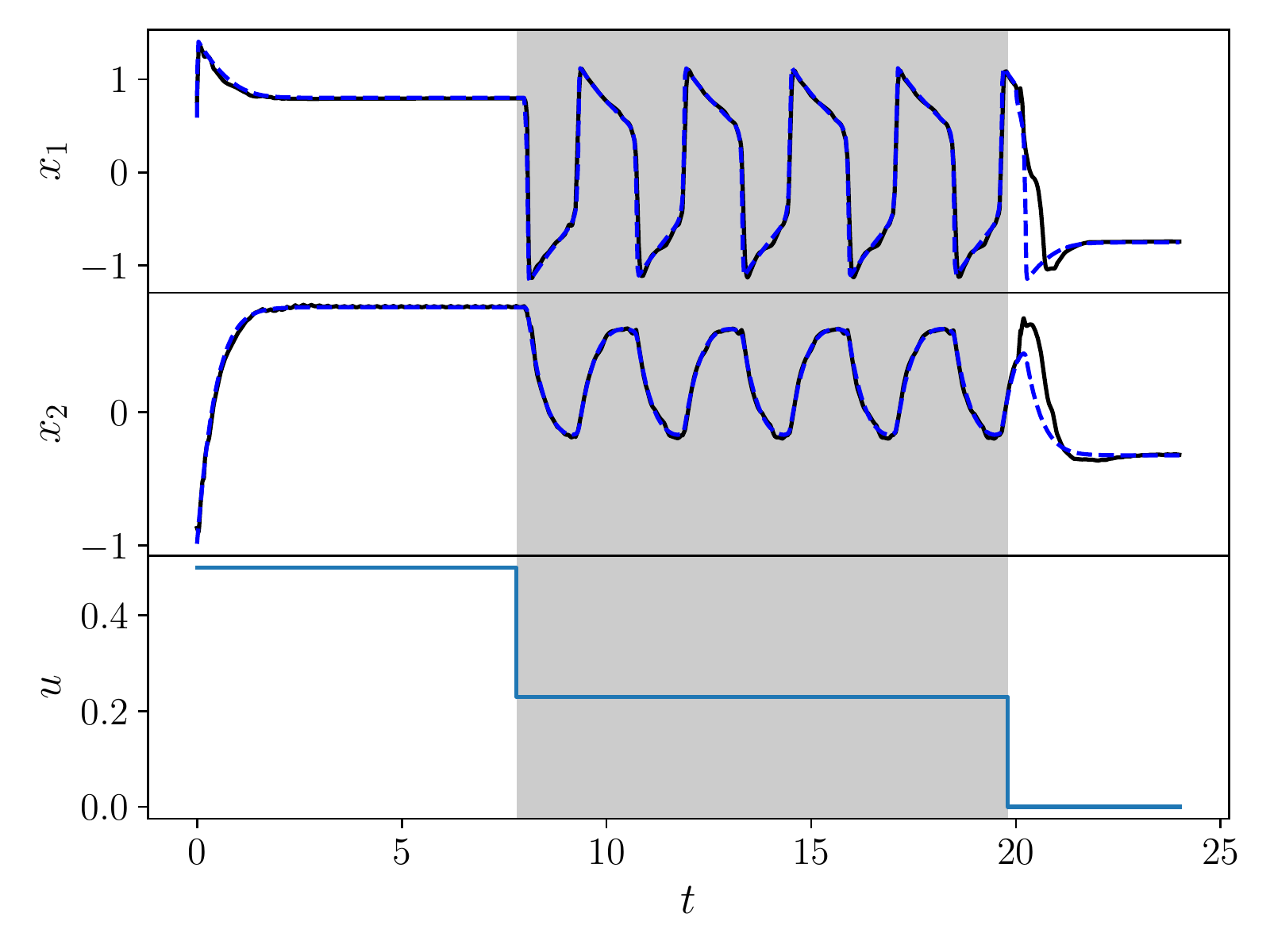}  
        \vspace{-0.5em}
        \subcaption{}
        \label{fig:fhn-excitability-a}
    \end{subfigure}
    \begin{subfigure}{.46\textwidth}
        \centering
        \includegraphics[width=.85\linewidth]{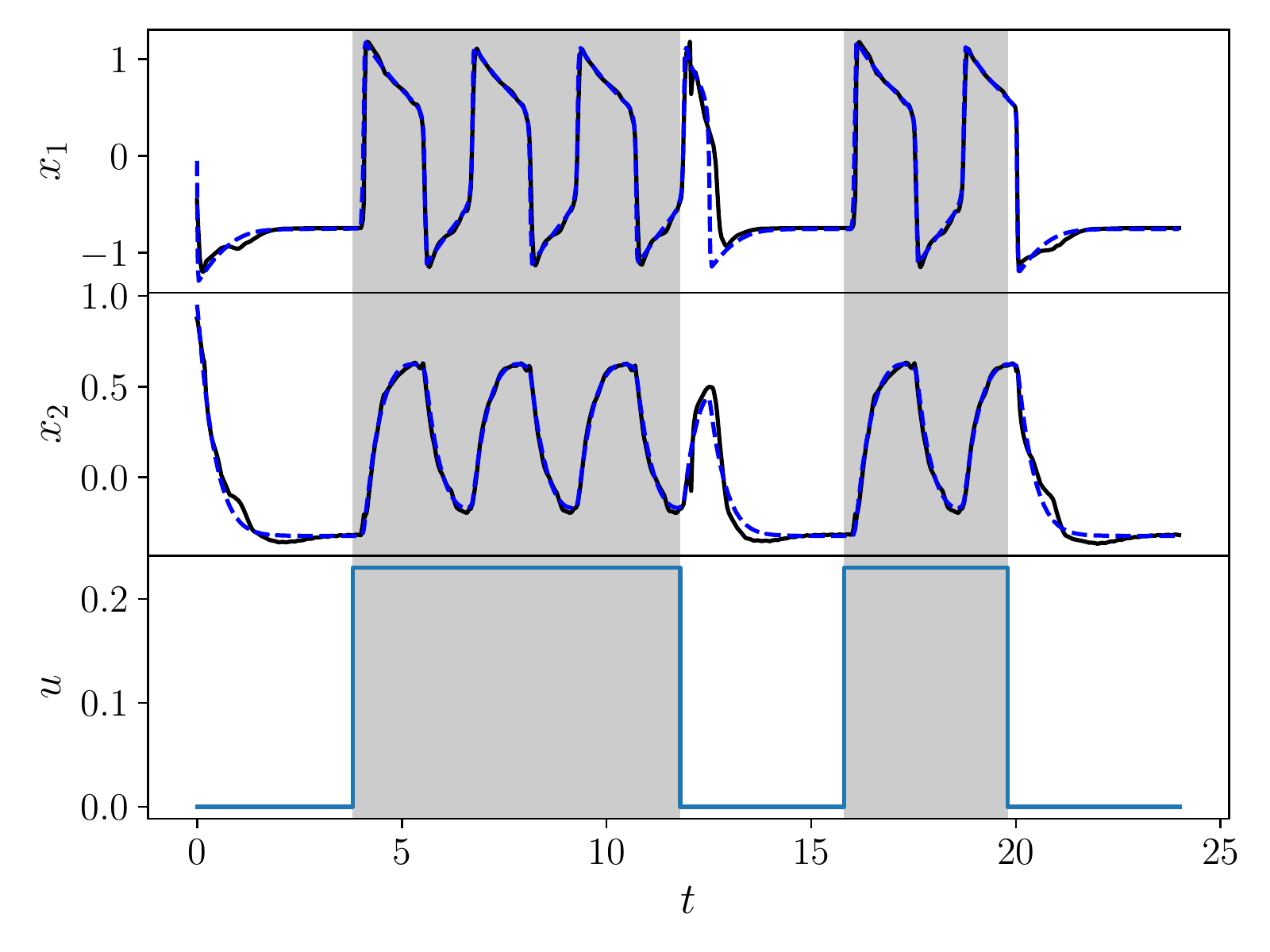} 
        \vspace{-0.5em}
        \subcaption{}
        \label{fig:fhn-excitability-b}
    \end{subfigure}
    \vspace{-0.5em}
    \caption{
       In (a) and (b), actual (blue, dashed) and predicted (black) trajectories demonstrate two distinct cases of excitability for the FitzHugh-Nagumo oscillator. Excitable regions are shaded in grey. 
    }
    \label{fig:fhn-excitability}
    \vspace{-0.5em}
        \hrulefill
\end{figure*}

Figure~\ref{fig:fhn-prediction} shows two predicted trajectories on $[0, 40]$ for two new pairs of
initial conditions and inputs drawn from $P_x$ and $P_u$ (i.e., unseen during training).
It can be observed (in Figure~\ref{fig:fhn-excitability}) that even though the trained model occasionally
fails to predict the peak value of the oscillations it subsequently recovers from the error.

\subsection{Prediction of excitability}
We further predict the occurrences of excitable behaviour in FitzHugh-Nagumo oscillator by carefully changing the amplitude of the input signal.
Note that, for the FitzHugh-Nagumo model~\eqref{eq:fhn-dynamics},
excitability is well-studied, allowing us to select the amplitude of the input suitable for exhibiting excitability.
In Figure~\ref{fig:fhn-excitability-a}, we show that the learned model is able to predict that,
as the amplitude of the input is gradually decreased over time,
the oscillator's response traverses from a higher resting potential (constant response),
passes through the excitable region (periodic spike train),
and returns to a lower resting potential (constant response).
In Figure~\ref{fig:fhn-excitability-b},
the model predicts the occurrence of two distinct excitable regions with two sets of spike trains.

\vspace{-1em}

\section{Concluding remarks}\label{sec:conclusion}
We presented a recurrent neural network architecture to learn the flow of a causal and time-invariant control system in continuous time from trajectory data.
Exploiting causality and time-invariance, we show that the problem of learning the flow function can be cast as the problem of learning a discrete-time dynamical system, motivating the use of an RNN-based architecture.
Our experimental results on the Van der Pol and FitzHugh-Nagumo oscillators show that the learned model has good prediction performance,
and demonstrate that the model is able to generalise to longer prediction time horizons and new classes of input signals.
We expect that our approach can provide an alternative to traditional modelling approaches in control problems,
bypassing the need of solving complex dynamics equations by directly predicting trajectories using the flow function.

There are many possible avenues of research to improve upon and develop the method we have proposed here.
An immediate extension would be the removal of the restriction to piecewise constant inputs, in order to obtain a more general class of continuous-time models.
Regarding further experimental validation, we intend to evaluate the method on other classes of systems.
A scalability study to systems with a large number of states would be particularly interesting.
It is also important to evaluate the model in the context of control~applications where specific forms of feedback are typically involved.
Theoretical properties of the problem of learning the flow function, as formulated in Section~\ref{sec:formulation}, have not been studied, to the best of our knowledge. It would be of interest to investigate whether it is possible to obtain error bounds or scaling
laws for the size of the optimal model.

Additionally, this work has significant implications for neuro-biological study as well as negative resistance circuits.
Learning the flow function can be viewed as a new method for in silico studies that is data-driven, independent of numerical integration and does not suffer from numerical instability.
For instance, learning excitable spiking patterns from an up-scaled network of neurons
(e.g., a network of Hodgkin-Huxley oscillators with multiple ion-channels)
will directly follow from the methods developed in this paper.
From the point of view of electrical circuits,
learning the flow function is equivalent to learning the conductance of a negative resistance circuit if we consider the input to be a current signal and the state trajectories to be branch voltages. 


\section{Acknowledgements}
The computations were enabled by resources provided by the Swedish National Infrastructure for Computing (SNIC) at Chalmers Centre for Computational Science and Engineering (C3SE) partially funded by the Swedish Research Council through grant agreement no.~2018-05973.

\bibliography{ifacconf.bib}
                                            
\end{document}